\documentclass[prb,showpacs,preprintnumbers,amsmath,amssymb,endfloats,letterpaper,footinbib]{revtex4}

\usepackage{graphicx}
\usepackage{dcolumn}
\usepackage{bm}
\usepackage{epstopdf}
\usepackage{latexsym}

\newcommand{\be}{\begin{equation}}
\newcommand{\ee}{\end{equation}}
\newcommand{\bea}{\begin{eqnarray}}
\newcommand{\eea}{\end{eqnarray}}

\newcommand{\req}[1]{Eq.~(\ref{#1})}

\begin{document}

\title{Density of States,  Entropy, and the Superconducting Pomeranchuk Effect in Pauli-Limited Al Films}

\author{X. S. Wu}
\altaffiliation[Present address: ]{School of Physics, Georgia Institute of Technology, Altanta, GA, 30332}
\author{P.W. Adams}
\affiliation{Department of Physics and Astronomy, Louisiana State
University, Baton Rouge, Louisiana 70803, USA}

\author{G. Catelani}
\affiliation{Laboratory of Atomic and Solid State Physics, Cornell
University, Ithaca, New York 14853, USA}

\date{\today}

\begin{abstract}

We present low temperature tunneling density of states measurements
of Pauli-limited Al films in which the Zeeman and orbital
contributions to the critical field are comparable. We show that
films in the thickness range of 6-7 nm  exhibit a reentrant
parallel critical field transition which is associated with a high
entropy superconducting phase, similar to the high entropy solid
phase of $^3$He responsible for the Pomeranchuk effect.  This phase
is characterized by an excess of states near the Fermi energy so
long as the parallel critical field transition remains second order.
Theoretical fits to the zero bias tunneling conductance are in good
agreement with the data well below the transition but theory
deviates significantly near the transition.  The discrepancy is a
consequence of the emergence of $e$-$e$ interaction correlations as
one enters the normal state.

\end{abstract}

\pacs{74.25.Dw,74.78.Db,74.50.+r}
\maketitle
\section{Introduction}
Phase transitions are ubiquitous in physics and continue to be a
large and important part of current condensed matter research.
Perhaps the most common example of a classical phase transition is
one in which a system undergoes a first-order transition from a
disordered phase to a phase exhibiting long range order, such as a
liquid to solid transition.  Because liquid has a higher entropy,
latent heat must be removed from the system before the solid phase
can form.  Though examples are exceedingly rare in nature, systems
which undergo a transition from a disordered phase to an ordered
phase of {\it higher} entropy exhibit a number of counterintuitive
and fascinating properties such as reentrance and negative latent
heats.  The most famous example is the liquid to solid transition
in $^3$He.  Subtle correlations between its spin and coordinate
degrees of freedom produce an anomalously low entropy in liquid
$^3$He.  Consequently, the entropy of solid $^3$He is higher than
that of the liquid \cite{Richardson1996} near the melting curve.  As
was first pointed out by Pomeranchuk,\cite{Pomeranchuk1950} the
resulting negative latent heat can be used to produce a cooling
effect if one converts liquid to solid by increasing the pressure.
This is commonly known as the Pomeranchuk effect and it was, in
fact, the method used to cool liquid $^3$He in the series of
experiments that led to the discovery of superfluid $^3$He.
\cite{Richardson1996}  The only other quantum system known to
exhibit a high entropy ordered phase is Pauli-limited
superconductivity.\cite{Fulde1973,Continentino2005}  In analogy to
$^3$He, the anomalous entropy is associated with the interplay
between electronic spin and orbital degrees of freedom where the
superconducting phase plays the role of the solid, the normal phase
the liquid, and parallel magnetic field the role of pressure.  In
the present paper, we address the possibility of realizing a
superconducting version of the Pomeranchuk effect.

Magnetic fields generally have a detrimental effect on
superconductors via two independent channels. The first is an
orbital effect associated with the fact that cyclotron motion is
incompatible with the formation of Cooper pairs and hence
superconductivity; for the vast majority of superconducting
systems the critical field transition is completely dominated by the
orbital response of the conduction electrons. If a magnetic field is
applied in the plane of a superconducting film whose thickness is
much less than the superconducting coherence length and whose
electron diffusivity is low,\cite{Fulde1973} then the critical
field transition is, in fact, mediated by the Pauli spin
polarization of the electrons: the system will undergo a well
defined first-order phase transition to the normal state when the
electron Zeeman splitting is of the order of the superconducting
gap.\cite{Clogston1962,Chandrasekhar1962} When realized in low atomic
mass superconductors such as Al and Be, this ``spin-paramagnetic''
(S-P) transition, as it is commonly known, exhibits a variety of
novel phases along with some very unusual dynamics.\cite{Fulde1973,
Wu1995a, Butko1999, Suzuki1984,Wu2005}  Here we present tunneling
measurements of the density of states (DoS) in Al films which are
about 2-3 times thicker than those typically used in S-P studies.\cite{Butko1999a}
In this marginally Pauli-limited regime, the
critical field behavior of the films can become reentrant.\cite{Fulde1967}
The reentrance occurs because the superconducting
state has higher entropy than the normal state. This excess entropy
is expected to be reflected in an enhancement of the Fermi energy
DoS {\it above} that of the normal state.\cite{Fulde1967,Suzuki1984,Alexander1985}

In a classic S-P geometry a planar magnetic field is
applied to a film of thickness much less than the Pippard coherence
length $\xi$. The relative importance of the orbital
response compared with the spin polarization can be quantified by
the dimensionless orbital depairing parameter \cite{Fulde1973}
\begin{equation}
c=\frac{D(ed)^2\Delta_o}{6\hbar\mu_{B}^{2}}
\label{Orbital}
\end{equation}
where $D$ is the electron diffusivity, $d$ is the film thickness,
$e$ is the electron charge, $\Delta_o$ the zero-temperature, zero-field gap energy and
$\mu_{B}$ the Bohr magneton. Fulde and coworkers\cite{Fulde1973}
showed that in the absence of spin-orbit coupling, the low
temperature critical field behavior would remain first-order for
$c\le1.66$, but interestingly, for (thicker) films with values in the range
$1.66<c<2.86$, the critical field behavior was predicted to be
reentrant.  Reentrance has indeed been observed in Al\cite{Suzuki1983}
and TiN\cite{Suzuki2000} films of comparable
thickness to those used in this study.  Tunneling DoS studies,
however, have been quite limited and though considerable theoretical
work has been done on the problem,\cite{Fulde1967,Maki1969} to the best of our
knowledge a quantitative comparison between theory and experiment was never made
in the regime where the DoS enhancement is present.
\section{Experimental Details}
A series of Al films with thicknesses ranging from 6 to 8 nm and
corresponding 100 mK normal state sheet resistances ranging from
$R=10$ to 20 $\Omega$ were fabricated by e-beam deposition of
99.999\% Al onto fire polished glass microscope slides held at 84 K.\cite{Wu2005}
The depositions were made at a rate of $\sim0.2$ nm/s
in a typical vacuum $P<3\times10^{-7}$ Torr.  After deposition, the
films were exposed to the atmosphere for 0.5-4 hours in order to
allow a thin native oxide layer to form. Then a 14-nm thick Al
counterelectrode was deposited onto the film with the oxide serving
as the tunneling barrier. The counterelectrode had a parallel
critical field of $\sim1.1$ T due to its relatively large thickness,
which is to be compared with $H_{c\parallel}\sim3$ T for the films.
The junction area was about 1 mm$\times$1 mm, while the junction
resistance ranged from 10-20 k$\Omega$ depending on exposure time
and other factors.  Only junctions with resistances much greater
than that of the films were used.  Measurements of resistance and
tunneling were carried out on an Oxford dilution refrigerator using
a standard ac four-probe technique. Magnetic fields of up to 9 T
were applied using a superconducting solenoid. A mechanical rotator
was employed to orient the sample \textit{in situ} with a precision
of $\sim0.1^{\circ}$.
\section{Results and Discussion}
\begin{figure}
\includegraphics[width=5.5in]{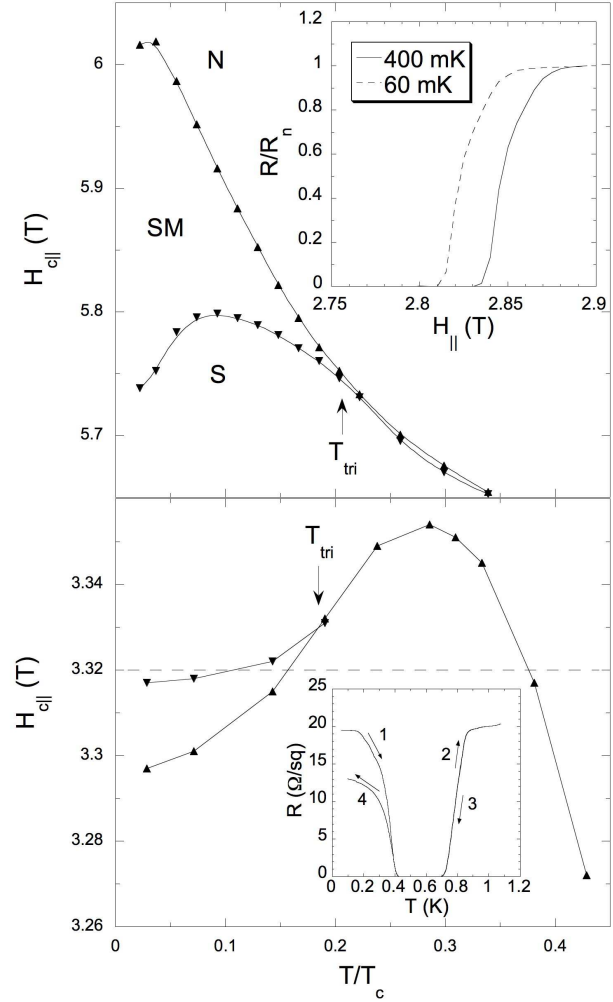}
\caption{\label{PD} Upper panel: phase diagram of a 2.5 nm thick Al
film ($T_c=2.7$ K) in parallel field, where S is the superconducting
phase, N the normal state, SM is the state memory coexistence
region, and $T_{tri}$ is the tricritical point.  Lower panel: phase
diagram of a 6 nm thick Al film ($T_c=1.9$ K).  Note the local
maximum in the second-order portion of the critical field curve. The
horizontal dashed line represents a reentrant cut through the phase
diagram. Upper inset:  Two resistive critical field transitions in the 6
nm film showing a higher $H_{c||}$ at higher temperature. Lower inset: Reentrant
superconducting transition corresponding to the cut through the 6 nm phase diagram represented by
the dashed line.  The numbers represent the temperature scan sequence.}
\end{figure}

Shown in the upper panel of Fig.~\ref{PD} is the temperature
dependence of the parallel critical field of a 2.5 nm thick Al film
having $T_c=2.7$ K, $\Delta_o/e=0.41$ mV, $\xi\sim15$ nm, and sheet
resistance $R\sim1$ k$\Omega$.   Using the bulk DoS for Al
$\nu_o=2\times10^{22}$ ${\rm(eVcm^3)}^{-1}$ to estimate $D$ in
\req{Orbital}, the orbital depairing parameter for this film is
$c=0.02$.\cite{SOcriticalfield} The upper panel of Fig.~\ref{PD}
represents the classic S-P phase diagram in which a high temperature
line of second order phase transitions terminates into a line of
first order transitions at the tricritical point $T_{tri}\sim0.3T_c$.\cite{Wu1994}
The low temperature superconducting (S) and normal
(N) phases are separated by a robust coexistence region in which the
state of the system is solely determined by the system state prior
to entering the region, i.e.\ the state memory region (SM).\cite{Butko1999}
In the lower panel of Fig.~\ref{PD} we show the
corresponding phase diagram for a 6 nm thick Al film having
$T_c=2.1$ K, $\Delta_o/e=0.32$ mV, $\xi=56$ nm, and $R=20\ \Omega$.
Note that this thicker film has a substantially lower critical field
as a consequence of orbital depairing, and that the structure of its
phase diagram is dramatically altered. The critical field behavior
is now reversibly reentrant by virtue of the local maximum in the
parallel critical field near $T/T_c=0.45$. Indeed, a naive estimate
of the orbital depairing parameter $c\sim 2.2$ falls within the
reentrant range. In the upper inset we show that the resistive parallel
critical field at 400 mK is higher than it is at 60 mK.   Reentrance
is easily demonstrated by cutting across the phase diagram along the
dashed line in Fig.~\ref{PD}, as is shown in the lower inset.  The
numbers in this curve represent the sequence of temperature scans at
a constant field. The reentrant region was entered at 60 mK by
lowering the field from a supercritical value of 3.5 T to the field
at which the temperature scans were performed, 3.320 T.  The low
temperature hysteresis is a consequence of the fact that the
transition becomes first order below $\sim0.4$ K .

\begin{figure}
\includegraphics[width=6in]{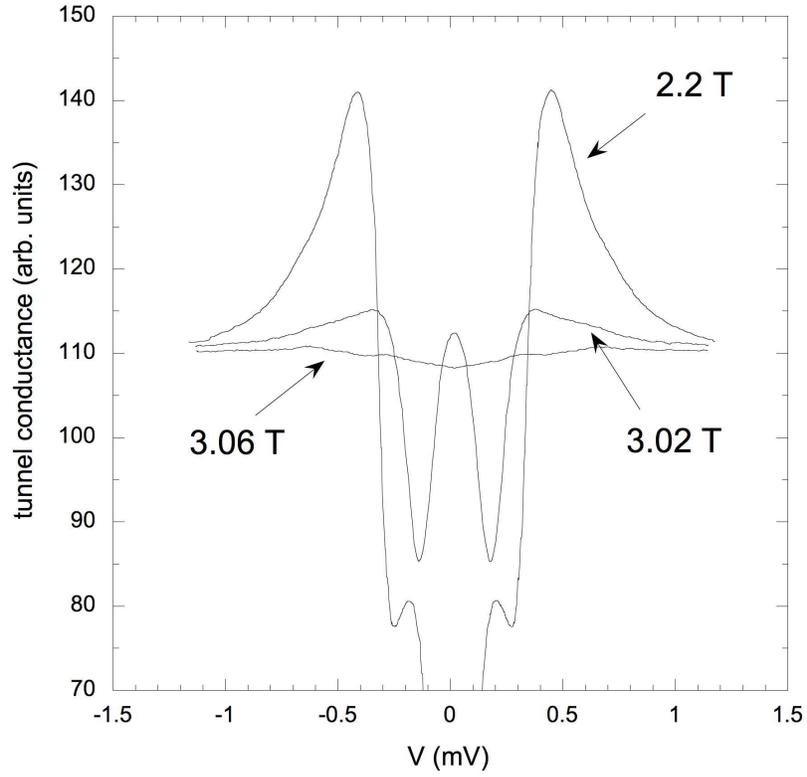}
\caption{\label{Spectra} Tunneling density of states spectra of a 7
nm Al film in several parallel magnetic fields.  Note that the zero
bias peak in the 3.02 T curve rises above the 3.06 T normal state
curve.  The modest suppression of the DoS at zero bias in the normal
state curve is an $e$-$e$ interaction effect and is commonly known
as the zero bias anomaly.}
\end{figure}

 Shown in Fig.~\ref{Spectra} are tunneling spectra at several values of parallel critical field.
At the temperatures at which these data were taken the tunneling conductance is simply proportional
to the electronic DoS.\cite{Tinkham1996} The Zeeman splitting of the BCS coherence peaks can be seen in the 2.2 T curve.\cite{Meservey1970} As the field is increased, the gap begins
to collapse due to orbital pair breaking, and near the critical field
transition the subgap Zeeman features reach the Fermi energy, thereby producing
a zero bias peak in the DoS curve.  This superconducting DoS
enhancement is evident in the 3.02 T curve of Fig.~\ref{Spectra},
where the DoS at V = 0 is clearly higher than that of the 3.06 T
normal state curve.

\begin{figure}
\includegraphics[width=6in]{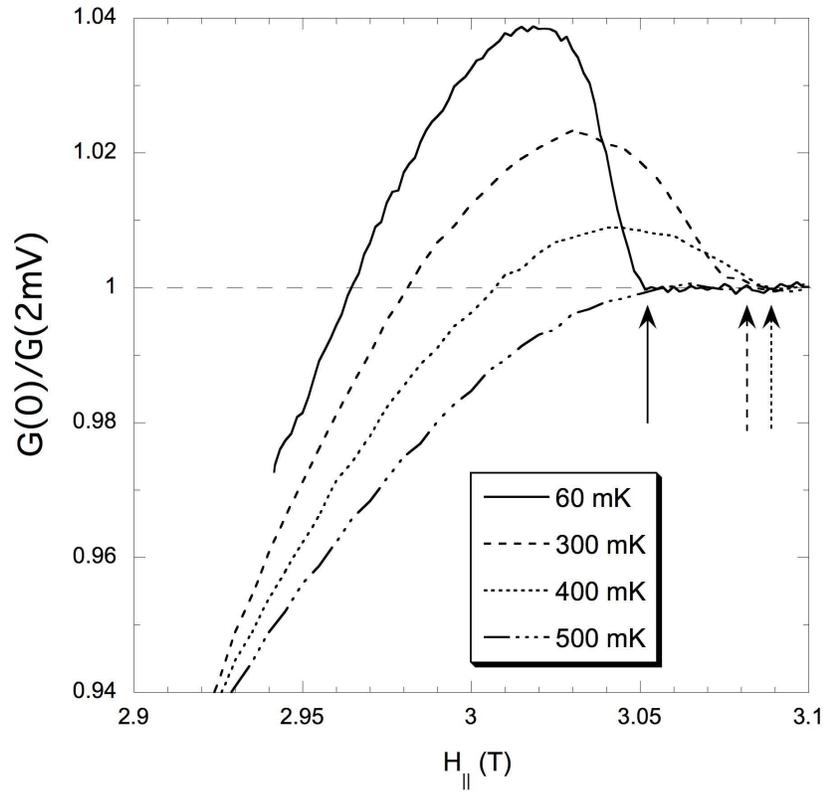}
\caption{\label{ZBspectra} Normalized density of states at zero bias for a 7 nm thick film.   The peaks correspond to excess states at the Fermi
energy associated with the onset of superconductivity.    Note that the onset critical fields, indicated by the arrows, are not monotonic in $T$. }
\end{figure}

\begin{figure}
\includegraphics[width=6in]{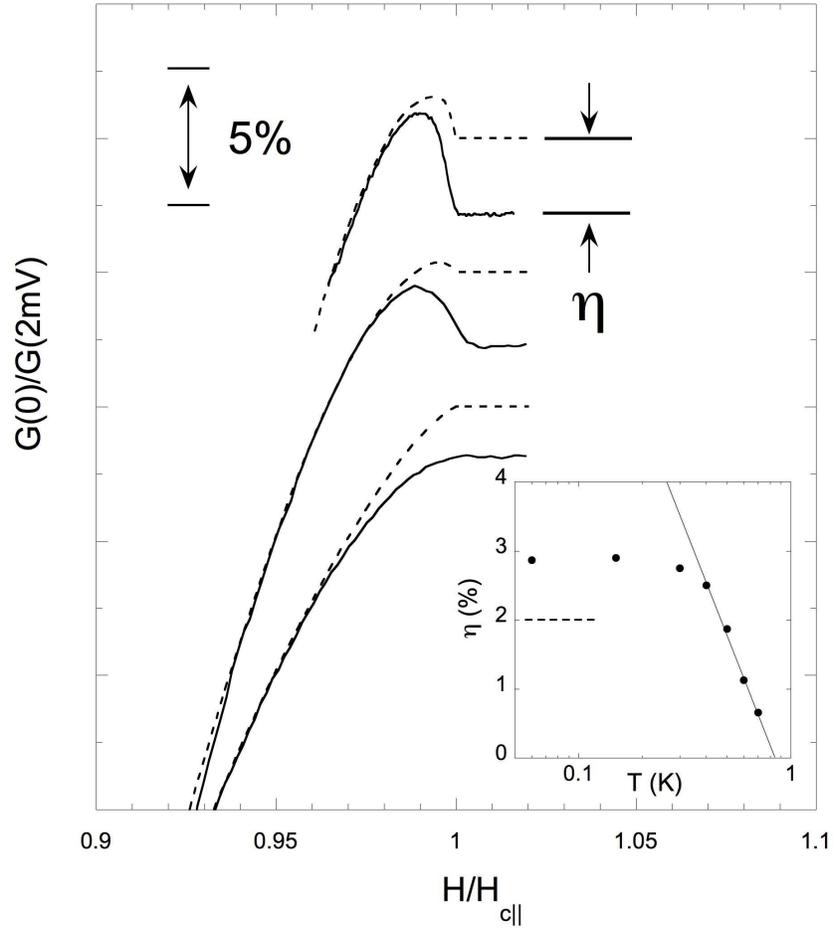}
\caption{\label{ZBfits}   The dashed lines represent
theoretical fits to the zero bias tunneling data at 60
mK, 300 mK, and 500 mK, top to bottom.  The respective curves have been shifted for
clarity.  Only that portion of the data well below the
critical field was used in the fits.  The magnitude of the
discrepancy between the theory and data near the transition is
represented by the parameter $\eta$. Inset: magnitude of the offset
between theory and data in the normal state as a function of
temperature for the 7 nm Al film.  The solid line is provided as a
guide to the eye.  The horizontal dashed line represents the
relative magnitude of the zero bias anomaly in the film.}
\end{figure}

In Fig.~\ref{ZBspectra} we plot the zero-bias tunneling conductance
as a function of parallel field at a variety of temperatures for a 7
nm Al film of resistance $R=15.8\ \Omega$.   The tunneling
conductance has been normalized by the normal state value.  The
arrows indicate the onset of superconductivity.  Because of the
reentrance effect, the onset is not monotonic in temperature.
Suzuki and coworkers\cite{Suzuki1983,Suzuki1987} reported a $\sim1\%$ DoS
enhancement over only a very narrow range of temperatures 550 mK $<T<$ 650 mK
in Al films in the thickness range of $7$-$9$ nm.
In contrast to these earlier reports, the DoS
enhancement peak appears below 500 mK and, in fact, grows in
magnitude with decreasing temperature. This film did not exhibit a
tricritical point, therefore we believe that the transition
remained second order down to 60 mK. Theoretically one can compute
such curves, assuming that the film's parameters are known, by
numerically solving the self-consistent equations for the order
parameter and the ``molecular'' magnetic field together with the
Usadel equations for the semiclassical Green's functions; these
equations are derived in Ref.~\onlinecite{Alexander1985}, and we refer the
reader to this work for the details (see also Ref.~\onlinecite{Suzuki2000} for
numerical calculations of the temperature dependence of the critical field).
Here we remind that the input parameters needed to
solve these mean-field equations at a given temperature $T$ are: the
orbital pair-breaking parameter $c$ (cf. \req{Orbital}); the spin-orbit scattering
parameter $b=\hbar/3\Delta_o \tau_{so}$, where $\tau_{so}$ is the
spin-orbit scattering time; the Fermi-liquid parameter
$G^0$ which, in the normal state, describes the renormalization of
the spin susceptibility; and the gap $\Delta_o$. In
Fig.~\ref{ZBfits} we compare our data to theoretical curves calculated using
parameters values $c=0.79$, $b=0.052$, $G^0=0.155$, and the measured gap
$\Delta_o =0.32$ meV, for three different temperatures. We note that the value
of $b$ is in agreement with what is found in the literature,\cite{Alexander1985}
while $G^0$ is consistent with what was previously measured in experiments
performed in the normal state on similar samples.\cite{Butko1999}
On the other hand, the value of $c$ used to fit the data is significantly lower than
that estimated by \req{Orbital} and ostensibly below the reentrant
threshold discussed above. However, the range of $c$ over which
reentrance can be observed is affected by both spin-orbit scattering
and the Fermi-liquid effect, as discussed in detail e.g. in Refs.~\onlinecite{Fulde1973,Suzuki2000}.
Furthermore, \req{Orbital} is valid
only in the ``local'' limit $\ell \ll d$, where $\ell$ is the mean
free path for impurity scattering. A more general relation for $c$
is found by multiplying \req{Orbital} by a function $f(\ell/d)$,
which describes the cross-over between the ``local'' and
``non-local'' ($\ell/d \gg 1$) electrodynamics in thin films,\cite{Maki1969}
and whose asymptotic behavior is $f(x)\simeq 3/4x$ for
$x\gg 1$; from the measured conductance of our sample, the bulk DoS
$\nu_o$, and the bulk Fermi energy $E_F = 11.7$eV, we estimate that
$\ell/d \approx 3.4$ and hence $c\simeq 0.73$, in good agreement
with the numerical fitting; therefore the values of all the three fitting
parameters $c$, $b$ and $G^0$ are consistent with independent measurements or estimates.

At fields well below the critical field there is excellent agreement
between theory and experiment. As the field increases, however, the
measured zero-bias tunneling DoS falls below the computed value by an amount
we define as $\eta$ in Fig.~\ref{ZBfits}. The origin of $\eta$ cannot
be attributed to superconducting fluctuations near the critical field, as
the Cooper pairing contributions to the DoS are shifted by the parallel magnetic
field to energies of the order of the Zeeman energy;\cite{Butko1999} it can, however,
be explained as follows: increasing the field depresses the order
parameter, so that more electrons can contribute to interaction
correlations which are not included in the mean-field analysis of
Ref.~\onlinecite{Alexander1985}; these correlations cause a
renormalization of the DoS close to the transition. In fact, it is
well known that  $e$-$e$ interactions lead to a logarithmic zero bias
correction to the 2D tunneling DoS, $\delta\nu \propto \ln (T\tau)$,
where $\tau$ is the elastic scattering time.\cite{Altshuler1985} In
the inset of Fig.~\ref{ZBfits} we plot $\eta$ as a function of $\ln
T$.  Note that above 300 mK, $\eta$ is well described by a $\ln T$
temperature dependence.  The low temperature saturation is partly a
consequence of the $\sim10$ $\mu$V probe voltage used to measure the
tunnel conductance.\cite{ZBAenergydep} The horizontal dashed line in the inset
represents the magnitude of the ZBA as obtained by a direct
measurement of the normal state DoS at 60mK.  Note that the ZBA and
$\eta$ are of the same magnitude and that a close inspection of the
data in Fig.~\ref{ZBfits} reveals that the region over which there
is disagreement between mean-field theory and experiment grows with
increasing temperature, as would be expected.

\begin{figure}
\includegraphics[width=6in]{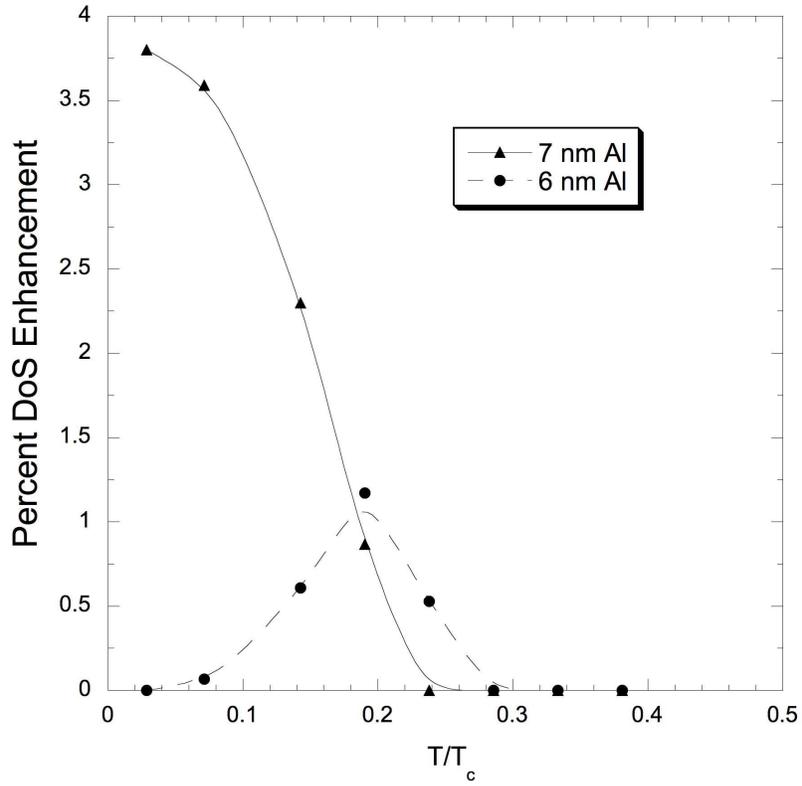}
\caption{\label{ZBmagnitude}Magnitude of DoS enhancement peaks such as those in
Fig.\ \ref{ZBspectra} as a function of reduced temperature for a 6
nm and a 7 nm thick Al film.  The 7 nm film did not exhibit a finite
tricritical point.  The 6 nm film critical field transition became
first-order below $T/T_c\sim0.2$. }
\end{figure}

In Fig.~\ref{ZBmagnitude} we plot the magnitude
of the DoS peak as a function of temperature for a 7 nm and a 6 nm
film.  Interestingly, the temperature dependence of the peak is
quite different for the two films.  The critical field behavior of
the 7 nm film did not show any signs of hysteresis down to 60 mK,
indicating that the transition remained second order.  In contrast,
the 6 nm film exhibited a tricritical point at $T/T_c\sim0.2$, as can
be seen in Fig.~\ref{PD}.  Note that the DoS peak in this film has
a local maximum at the tricritical point, and that the DoS
enhancement is rapidly extinguished as the temperature is lowered
below $T_{tri}$.  This may explain why earlier reports of the DoS
enhancements by Suzuki and coworkers\cite{Suzuki1984} were limited
to a relatively narrow range of temperatures, i.e. temperatures lying
between a tricritical point and the maximum of the critical field
curve.

In summary, we show not only that marginally thick Al films exhibit
reentrance, but also that the associated DoS enhancement can be
observed down to mK temperatures so long as the critical field
transition remains second order.  Since the Pomeranchuk effect is
based on the latent heat of a first order transition, it seems
unlikely that one can produce cooling in Al films.  Nevertheless,
the fact that the DoS enhancement peak occurs at the Fermi energy
affords one an unparalleled opportunity to study the interplay
between Coulombic $e$-$e$ interaction fluctuations and pairing
fluctuations in the vicinity of the transition.  Indeed, it is
fortuitous that the suppression of the DoS associated with the
former is comparable in magnitude to the enhancement of the latter.
\acknowledgments
We gratefully acknowledge enlightening discussions with Ilya Vekhter, Dana
Browne, and David Young. G.C. acknowledges financial support through the J.A. Krumhansl
Postdoctoral Fellowship.This work was supported by the National
Science Foundation under Grant DMR 02-04871.

\bibliographystyle{apsrev}
\bibliography{SPTmaster.bib}

\begin{thebibliography}{23}
\bibitem[{SOc()}]{SOcriticalfield}
\bibinfo{note}{All estimates are obtained by treating the films as quasi
  two-dimensional systems; this is justified, since in all cases the coherence
  lengths and thermal lengths are larger that the films' thicknesses.}
\end{thebibliography}

\newpage

\end{document}